\documentclass[prl,aps,twocolumn,preprintnumbers,amsmath,amssymb,nofootinbib,superscriptaddress,notitlepage]{revtex4-1}
\usepackage{epsfig}
\usepackage[utf8]{inputenc}
\usepackage{color}
\usepackage{epstopdf}
\usepackage{multirow}
\usepackage{verbatim}
\usepackage{ftnxtra}
\usepackage{slashed}
\usepackage[normalem]{ulem}
\usepackage{subfigure}
\usepackage{booktabs}
\usepackage{tikz}
\usepackage{braket}

\newcommand{\LG}{\text{LG}}

\newcommand{\RNum}[1]{\uppercase\expandafter{\romannumeral #1\relax}}

\newcommand{\Per}{\text{Per}}

\newcommand{\uestc}{\affiliation{School of Physics, University of Electronic Science and
Technology of China, Chengdu 610054, China}}
\newcommand{\ucas}{\affiliation{School of Physical Sciences, University of Chinese Academy of Sciences (UCAS), Beijing 100049, China}}
\newcommand{\kek}{\affiliation{KEK Theory Center, Institute of Particle and Nuclear Studies (IPNS), High Energy Accelerator Research Organization (KEK), 1-1 Oho, Tsukuba, Ibaraki, 305-0801, Japan}}

\newcommand{\cssm}{\affiliation{CSSM, Department of Physics, Adelaide University, Adelaide, 5005, South Australia, Australia}}

\begin{document}

\title{
General Hamiltonian Approach to the $\mathbf{N}$-Body Finite-Volume Formalism:\\ Extracting the $\mathbf{\omega}$ Resonance Parameters from Lattice QCD
}

\author{Kang Yu}\email{yukang21@mails.ucas.ac.cn}
\ucas

\author{Derek B. Leinweber}\email{derek.leinweber@adelaide.edu.au}
\cssm

\author{Anthony W. Thomas}\email{anthony.thomas@adelaide.edu.au}
\cssm

\author{Guang-Juan Wang}\email{wgj@post.kek.jp}
\kek

\author{Jia-Jun Wu}\email{wujiajun@ucas.ac.cn, corresponding author}
\ucas

\author{Zhi Yang}\email{zhiyang@uestc.edu.cn}
\uestc

\preprint{KEK-TH-2817, ADP-26-05/T1302}

\date{\today}

\begin{abstract}

We present a nonperturbative Hamiltonian framework (NPHF) to address the general $N$-body problem. This framework rigorously connects finite-volume spectra from lattice QCD to scattering observables from experiment. 
To demonstrate its applicability, we extract the resonance parameters of the $\omega$ meson by simultaneously analyzing the isoscalar $3\pi$ and isovector $2\pi$ systems.
The Hamiltonian unifies single-particle $\omega$, two-particle $\rho\pi$, and three-particle $\pi\pi\pi$ dynamics within a single unitary formalism.
Using leading lattice QCD spectra from the Chinese Lattice QCD Collaboration at $m_\pi$ = 208 and 305 MeV, we perform a fit in the isovector and isoscalar channels, accurately describe the lattice spectra and obtain robust determinations of the $\rho$ and $\omega$ pole positions.
This work establishes a foundational approach for extracting resonance dynamics from finite-volume spectra. Given the ubiquity of three-body dynamics in exotic hadrons, halo nuclei, and neutron star matter, this general formalism holds broad relevance across particle, nuclear, and astrophysical physics.

\end{abstract}

\maketitle

Quantum systems with three or more interacting constituents pose a profound and persistent challenge across particle, nuclear, and astrophysical physics. 
Unlike the two-body problem, the $N\geq 3$ few-body problem is inherently non-separable. 
Its dynamics are governed by coupled equations that cannot be reduced to pairwise interactions without violating unitarity or analyticity.

From light nuclei to neutron stars, three-body dynamics play a decisive role. 
The dynamics are indispensable for describing the saturation of nuclear matter, the structure of Borromean halo nuclei such as $^6$He and $^{11}$Li, and reactions like $nd$ scattering, $(d,p)$ transfer, and three-body breakup processes~\cite{Greene:2017cik}. 
The triple-\(\alpha\) reaction (\(3\alpha \to {}^{12}\)C) relies critically on a resonant three-body state (the Hoyle state)~\cite{Epelbaum:2011md}, while the high-density equation of state of neutron stars is strongly sensitive to three-nucleon interactions constrained by multi-messenger observations~\cite{Hebeler:2013nza}.

In hadron physics, many resonances with masses above $1$ GeV$/c^2$, such as the $\pi(1300)$, $a_1(1260)$, $\eta(1405/1475)$, as well as exotic states like the $T_{cc}$, exhibit strong couplings to three-body channels. 
Subchannel resonances, rescattering effects, and triangle singularities can dramatically distort line shapes, shift pole positions, and bias extracted couplings if approximated within a two-body framework.
The Roper resonance exemplifies this complexity. 
It has often been modeled via effective channels like $N\sigma$ and $\Delta\pi$, in addition to $N\pi$ and bare nucleon states. 
However, the instability of both the $\sigma$ and $\Delta$ resonances causes relevant channels to decay directly into the physical $N\pi\pi$ continuum~\cite{Roper:1964zza}. 
These features underscore that a quantitatively reliable description of hadron structure demands a fully unitary treatment of three-body dynamics.

Lattice QCD (LQCD) has achieved remarkable success in computing the spectrum of stable hadrons, with results in excellent agreement with experiment~\cite{BMW:2008jgk,Fodor:2012gf,Dowdall:2012ab}. 
Recent advances now enable direct calculations of three-hadron correlation functions, including studies of the binding energy of ${}^3\text{He}$~\cite{NPLQCD:2012mex}, maximal-isospin systems (\(\pi^+\pi^+\pi^+\), \(K^+K^+K^+\))~\cite{Blanton:2019vdk,Fischer:2020jzp,Alexandru:2020xqf} and resonant channels involving the \(\omega\)~\cite{Yan:2024gwp}, \(\pi(1300)\)~\cite{Yan:2025mdm}, and \(a_1(1260)\)~\cite{Mai:2021nul}. 
However, extracting physical resonance parameters from these finite-volume spectra requires a formalism that rigorously incorporates three-body unitarity, which is beyond the scope of traditional two-body methods. 
%
%
We also anticipate that finite-volume effects in multi-hadron systems ($N \geq 4$) will inevitably need to be addressed.

Considerable effort has been devoted to developing formalisms with which to handle finite-volume effects in three-body systems~\cite{Polejaeva:2012ut, Hansen:2014eka, Hansen:2015zga, Briceno:2017tce, Guo:2017crd, Hammer:2017uqm, Hammer:2017kqx, Mai:2017bge, Doring:2018xxx, Briceno:2018aml, Briceno:2018mlh, Guo:2018ibd, Blanton:2019igq, Pang:2019dfe, Romero-Lopez:2019qrt, Blanton:2020gmf, Blanton:2020jnm,Romero-Lopez:2020rdq, Muller:2020wjo, Muller:2020vtt, Muller:2021uur, Muller:2022oyw, Jackura:2022gib, Baeza-Ballesteros:2023oph, Hansen:2020zhy, Pang:2020pkl, Blanton:2020gha, Hansen:2021sog, Blanton:2021mih, Blanton:2021llb, Severt:2022sku, Draper:2023xvu, Bubna:2023qnq, Briceno:2025yuq}.
Three distinct but conceptually equivalent~\cite{Jackura:2019bmu} formalisms have emerged: Relativistic Field Theory (RFT)~\cite{Hansen:2014eka,Hansen:2015zga}, Non-Relativistic Effective Field Theory (NREFT)~\cite{Hammer:2017kms,Hammer:2017uqm} and Finite Volume Unitarity (FVU)~\cite{Mai:2017bge}, as reviewed in Refs.~\cite{Hansen:2019nir,Mai:2021lwb}.
These frameworks establish a direct relation between the two-body scattering $T$-matrices, the low-energy effective constants of three-body interactions and the three-body finite volume energy spectrum. 
However, extracting resonance pole positions still requires an intermediate parameterization of the scattering amplitudes. 
For instance, in the recent lattice study of the $\omega$ meson~\cite{Yan:2024gwp}, two distinct ans\"atze—the so-called Generic method and an effective field theory—were employed to model the three-body dynamics and infer the pole location. 
Any model dependence necessarily introduces ambiguity in the extracted resonance properties.

In this Letter, we introduce a fundamentally different approach based on the finite-volume Hamiltonian formalism originally proposed in 2013~\cite{Hall:2013qba}. 
This nonperturbative Hamiltonian framework (NPHF) can address the general $N$-body problem relating finite-volume spectra to the scattering observables of experiment.  
Here we illustrate the framework through an explicit examination of finite-volume effects in the three-body system of the $\omega$ meson and their relation to its resonance properties.

The original finite-volume Hamiltonian approach \cite{Hall:2013qba} is now well-established for multi-channel two-body systems~\cite{Yu:2025gzg}. 
Over more than a decade of development, it has been successfully applied to various hadron systems, such as the $\mathrm{N}^*(1535)$~\cite{Liu:2015ktc}, $\mathrm{N}^*(1650)$~\cite{Abell:2023nex}, $\mathrm{N}^*(1440)$~\cite{Wu:2017qve,Owa:2025mep}, $\Lambda^*(1405)$~\cite{Hall:2014uca,Liu:2016wxq}, the positive parity $D_s$ meson family~\cite{Yang:2021tvc}, as well as the $\rho$ meson system~\cite{Wang:2025hew,Allton:2005fb}. 

The fundamental channels of this method can involve different particle numbers. 
For example, the calculation of $D_{s0}(2317)$ necessarily involves the bare state predicted in the Godfrey-Isgur model as well as the $DK$ channel~\cite{Yang:2021tvc}.
This framework directly connects the finite volume spectra ({\it i.e.}, lattice spectrum data), $S$-matrix parameters ({\it i.e.}, experimentally measured phase shifts and inelastic coefficients), and $T$-matrix complex-plane poles ({\it i.e.}, bound states, resonant states, or virtual states).
It is consistent with two-body short-range interactions and the usual L\"uscher formula within exponentially suppressed errors~\cite{wu:2014vma, Yu:2025gzg} and it can go beyond the traditional L\"uscher formula to successfully handle long-range interactions~\cite{Meng:2021uhz,Yu:2025gzg}. 
Therefore, extending the Hamiltonian approach, which has proven highly effective for two-body systems, to three-body systems is undoubtedly an important endeavor.

While the extension to three-body systems via the addition of coupled channels is conceptually natural, it presents significant technical challenges. 
First of all, the dimension of the three-body Hamiltonian matrix in momentum space becomes extremely large.
Furthermore, the Hamiltonian matrix needs to be block-diagonalized regarding the lattice symmetry group and isospin flavor symmetry to compare with lattice spectra.
In addition, in two-body subsystems, the rest and moving frames are mixed.
Finally, the scattering equation in the infinite volume is more difficult to solve.
This Letter addresses these challenges, demonstrating the solution of three-body finite volume eigenvalues within the NPHF as it is applied to the $\omega$ system.

The Hamiltonian basis states include single-particle states (bare states) and multi-particle channels. 
In the continuum momentum space, the bare state $\ket{B,\sigma}$ is a single state with the bare mass $m_B$ and its polarization $\sigma$, while the $N$-particle state is denoted by $\ket{\alpha^{(N)}_a(\vec{k}_1,\sigma_1;\cdots;\vec{k}_N,\sigma_N)}\equiv\ket{\alpha_a(\{\vec{k},\,\sigma\}_N)}$.
To incorporate isospin symmetry from the outset, we let $\alpha$ collectively denote the channel, isospin quantum number $(I,I_z)$ and irreducible representation (irrep) of the permutation group $S_N$, characterized by partition number  $[\kappa]$\footnote{We assume all $N$ particles belong to the same isospin multiplet (e.g., $3\pi$ or $3$N). The case where only $M<N$ particles belong to the multiplet needs minor modifications.}.
The index $a=1, ..., \text{dim}[\kappa]$ specifies the column of $[\kappa]$. 
The normalizations of the basis states are defined as
\begin{align}
&\bra{B',\sigma'}B,\sigma\,\rangle  =\delta_{B'B}\delta_{\sigma'\sigma},
\\
&\bra{\alpha^{(M)}_{a}(\vec{k}_1'\sigma_1',\cdots,\vec{k}_M'\sigma_M')}\beta^{(N)}_{b}(\vec{k}_1\sigma_1,\cdots,\vec{k}_N\sigma_N)\rangle
=\notag \\
&\delta_{MN}\delta_{\alpha\beta}\sum\limits_{s\in S_N}\left(\prod\limits_{i=1}^N \delta^3(\vec{k}'_{s_i}-\vec{k}_i)\delta_{\sigma'_{s_i}\sigma_i}\right) \mathcal{R}_{ab}^{[\kappa]}(s)\delta(s)\, ,
\label{eq:norInf}
\end{align}
where $\mathcal{R}^{[\kappa]}$ is the matrix representation of the permutation $s\in S_N$ irrep $[\kappa]$, with $(1,\cdots,N)\overset{s}{\to}(s_1,\cdots,s_N)$. 
Here $\delta(s)=\mathcal{R}^{[N]}(s)$ for bosonic systems and $\mathcal{R}^{[1^N]}(s)$ for fermionic systems.

The Hamiltonian is decomposed into a free energy part, $\hat{H}_0$, and an interaction part, $\hat{V}$.
$\hat{H}_0$ is given by
\begin{align}
    \hat{H}_0 \ket{\alpha_a(\{\vec{k},\,\sigma\}_N)} = \omega_{\alpha}(\vec{k}_1,\cdots,\vec{k}_N) \ket{\alpha_a(\{\vec{k},\,\sigma\}_N)},
\end{align}
with  $\omega_\alpha(\vec{k}_1,\cdots,\vec{k}_N)=\sum\limits_{i=1}^N \sqrt{ m_{\alpha_i}^2 + \vec{k}_i^2}$. 
$\hat{V}$ is given by
\begin{align}
& \bra{B}\hat{V}\ket{\alpha_a(\{\vec{k},\,\sigma\}_N)} \equiv  V^{B\alpha}_a(\{\vec{k},\,\sigma\}_N),
\\
&\bra{\alpha'_b(\{\vec{k}',\,\sigma'\}_M)}
\hat{V}\ket{\alpha_a(\{\vec{k},\,\sigma\}_N)
}
\nonumber \\
&\qquad\qquad\qquad\qquad  \equiv V^{\alpha'\alpha}_{ba}(\{\vec{k}',\,\sigma'\}_M,\{\vec{k},\,\sigma\}_N).
\end{align}
$V^{B\alpha}$ ($V^{\alpha'\alpha}$) represents the potential between bare states $B$ (multi-particle state $\alpha'$) and multi-particle states $\alpha$. 
These potentials are parameterized using effective field theory or phenomenological models.
Lattice calculations typically impose periodic boundary conditions, restricting momenta to discretized modes $\vec{k}_{\vec{n}}=2\pi\vec{n}/L$ with $\vec{n}\in\mathbb{Z}^3$ and volume extension $L$. 
The corresponding finite-volume multi-particle state is denoted as 
$\ket{\alpha_{a,L}(\{\vec{n},\,\sigma\}_M)}$, with the normalization identical to the infinite-volume case in Eq.~(\ref{eq:norInf}) except for the replacement $\delta^3(\vec{k}_i'-\vec{k}_j)\to\delta_{\vec{n}_i',\vec{n}_j}$. 
While the free energy part of the Hamiltonian remains unchanged in the finite volume, the interaction part ($\hat{V}^\text{F}$) needs an additional Fourier factor reflecting the conversion from continuum to discrete momentum states,
\begin{align}
&\frac{\bra{B}\hat{V}^\text{F}\ket{\alpha_{a,L}(\{\vec{n},\,\sigma\}_M)}}{\bra{B}\hat{V}\ket{\alpha_a(\{\vec{k}_{\vec{n}},\,\sigma\}_M)}}=\left(\frac{2\pi}{L}\right)^\frac{3(M-1)}{2}, \label{eq:vfinBC}\\
&\frac{\bra{\beta_{b,L}(\{\vec{n}',\,\sigma'\})_M}\hat{V}^\text{F}\ket{\alpha_{a,L}(\{\vec{n},\,\sigma\}_N)}}
{\bra{\beta_{b}(\{\vec{k}_{\vec{n}'},\,\sigma'\}_M)}\hat{V}\ket{\alpha_{a}(\{\vec{k}_{\vec{n}},\,\sigma\}_N)}}=
\left(\frac{2\pi}{L}\right)^\frac{3(M+N-2)}{2}.\label{eq:vfinCC}
\end{align}

The lattice spectrum is identified with the eigenvalues of the finite volume Hamiltonian $\hat{H}^\text{F}$. 
However, obtaining these eigenvalues meets a computational challenge due to the increased dimension of the matrix, particularly for $N \ge 3$. 
For example, with $\sum_{i}^N\vec{n}_i=\vec{0}$ and $|\vec{n}_i|\leq20$, there are about $400$ distinct $(\vec{n}_1,\vec{n}_2)$ for $N=2$. 
This increases to roughly $70000$ distinct $(\vec{n}_1,\vec{n}_2,\vec{n}_3)$ for $N=3$.

In practice, lattice spectra are usually extracted from interpolating fields that have been projected onto a certain irrep of the lattice symmetry group $G$ in the direct product space of momentum and total isospin.
Consequently, $\hat{H}^\text{F}$ must be block-diagonalized. 
This is achieved as follows.

In the first step, the $\sigma$-polarization representation is converted into the $\lambda$-helicity representation, in which $\lambda$ remains unchanged under proper rotation and changes sign under parity inversion. 
We then select a set of reference momenta ${N}^{r}=\{\{\vec{n}^r\}_1,\{\vec{n}^r\}_2,\cdots\}$ such that the discretized momentum space decomposes into the direct sum of their $(G\times S_N)$-orbit. 
%
As a result, the Hilbert space, $\mathcal{H}$, can be decomposed as the direct sum of several $G\times S_N$-invariant subspaces:
\begin{align}
\mathcal{H} &= \bigoplus\limits_{\alpha}\bigoplus\limits_{\{\vec{n}^r\}\in{N}^r}\bigoplus\limits_{\{\lambda\}} \,\mathcal{S}_{\alpha}(\{\vec{n}^r,\,\lambda\}_N) \equiv \bigoplus\limits_\xi \mathcal{S}_\xi \, ,
\\
\mathcal{S}_\xi& \equiv \text{span}\left\{ \hat{g}\ket{ \alpha_a(\{\vec{n}^r,\,\lambda\}_N) } \,|\,g \in G\,,a=1,\cdots\right\} \, ,
\end{align}
where the collective notation $\xi=(\alpha,\{\vec{n}^r,\,\lambda\}_N)$ characterizes the $G\times S_N$-invariant subspace. 
It should be noted that the range of $\{\lambda\}$ depends on the given $\{\vec{n}^r\}$. 
The key point is that, two $\{\lambda\}$s may be equivalent, in the sense that the two corresponding $\mathcal{S}_\xi$ are identical.
For the two-particle system in the rest frame, as an example, only one of $(\lambda_1,\lambda_2)$ and $(-\lambda_2,-\lambda_1)$ can be in the range since under parity inversion $\mathbb{P}\ket{\alpha_a(\vec{n}_1\lambda_1,\vec{n}_2\lambda_2)} \propto  \ket{\alpha_a(\vec{n}_2-\lambda_1,\vec{n}_1-\lambda_2)} \propto \ket{\alpha_a(\vec{n}_1-\lambda_2,\vec{n}_2-\lambda_1)}$. 
A general discussion is provided in the Supplemental Material.

The second step is to construct the orthonormal basis for each $\mathcal{S}_\xi$ that furnishes a definite irrep $\Gamma$ of $G$.
For a given $\xi$, we denote the desired basis states by $\ket{\Gamma,r,\mu}$ where $\mu$ and $r$ specify the column and the occurrence, respectively. 
A natural starting point is the set of the projected states  $\{
\sum_{g\in G} D^{\Gamma*}_{\mu\nu}(g)\, \hat{g}\,\ket{\alpha_a(\{\vec{n}^r,\,\lambda\}_N)}\}$ with $\nu=1, ..., \text{dim}[\Gamma]$ and $a=1, ...., \text{dim}[\kappa]$, respectively. 
However, these states are not linearly independent.
Their linear dependence relations are fully characterized by the following idempotent overlap matrix
\begin{align}
    &I_{b\nu^\prime,a\nu}= \nonumber
    \\
    & \sum\limits_{g\in \Per(\{\vec{n}^r\})}\!\!\!\! \langle \alpha_b(\{\vec{n}^r,\,\lambda\}_N) |\,\hat{g}\,| \alpha_a(\{\vec{n}^r,\,\lambda\}_N) \rangle\, {D}^*_{\nu^\prime\nu}(g) \, ,
    \label{eq:Imat}
\end{align}
where $\Per(\{\vec{n}\})$ is the subgroup of $G$ consisting of $g$ such that $(g\vec{n}_1,\cdots,g\vec{n}_N)$ is a permutation of $(\vec{n}_1,\cdots,\vec{n}_n)$. 
It is obvious that the little group $\LG(\{\vec{n}^r\})$, corresponding to the identity permutation, is its subgroup. 
Eq.~(\ref{eq:Imat}) can be further simplified with the explicit expression depending on $\xi$. 
It can be easily proven that the eigenvalues of the the $I$-matrix are either $0$ or a constant related to the order of $\Per(\{\vec{n}^r\})$ and $\{\lambda\}$. 
For each nonzero eigenvalue of $I$-matrix, the associated eigenvectors $c^r$ form an orthonormal basis that is both irreducible under the symmetry group and linearly independent.
\begin{align}\label{eq:ireep basis}
    \ket{\Gamma,r,\mu} &= \mathcal{N} \sum\limits_{b,\nu}\sum\limits_{g\in G} (c^r)_{b\nu}\, D^{\Gamma *}_{\mu\nu}(g)\,\hat{g}\, \ket{\alpha_b(\{\vec{n}^r,\,\lambda\}_N)}
    \notag \\
    &\equiv  \sum\limits_{\theta} \ket{\theta} \,\left[X^\Gamma_\xi\right]_{\theta,\,r\mu} \, ,
\end{align}
where $\mathcal{N}$ is a normalization factor that can be determined analytically. 
%
%
Since it is more often to parameterize the effective Hamiltonian in $\sigma$-polarization representation, we expand the irrep basis in terms of $\sigma$-states in the second line of Eq.~(\ref{eq:ireep basis}) and use the collective notation $\theta=(a,\{\vec{n},\sigma\}_N)$. 
The dummy indices $a$ and $\{\sigma\}$ run over all possible values, while $\{\vec{n}\}$ runs over the quotient space ($G$-orbit of $\{\vec{n}^r\}$)/$\sim$ with the equivalence relation $\{\vec{n}\}\sim\{\vec{n}'\}$ if they are related by permutation. 
Because of the Wigner-Eckart theorem, the physical result is $\mu$-independent so we can always take $\mu=1$ and suppress it.

Expanded in terms of $\ket{\xi;\,\Gamma,r}$, the interaction part of the projected Hamiltonian, $[{V}^\Gamma]$, can be built in a block-matrix manner
\begin{align}
    {V}^\Gamma =  
    \begin{pmatrix}
        [V^\Gamma_{\xi_1,\xi_1}] &  \cdots & [V^\Gamma_{\xi_1,\xi_i}] \\
        \vdots & \ddots &  \vdots \\
        [V^\Gamma_{\xi_i,\xi_1}] & \cdots & [V^\Gamma_{\xi_i,\xi_i}]
    \end{pmatrix},
\end{align}
where $[V^\Gamma_{\xi'\xi}]$ is the matrix given by similarity transformation $[V^\Gamma_{\xi'\xi}] = [X^\Gamma_{\xi'}]^\dagger \cdot [V^\mathrm{F}_{\xi'\xi}] \cdot [X^\Gamma_\xi]$ with the matrix $[V^\mathrm{F}_{\xi'\xi}] : \mathcal{S}_\xi \to \mathcal{S}_{\xi'}$ of the form
\begin{align}
    [V^\mathrm{F}_{\xi' \xi}] = 
    \begin{pmatrix}
        [V^\mathrm{F}]_{\theta'_1,\theta_1} &  \cdots & [V^\mathrm{F}]_{\theta'_1,\theta_j} \\
        \vdots & \ddots &  \vdots \\
        [V^\mathrm{F}]_{\theta'_{j'},\theta_1} & \cdots & [V^\mathrm{F}]_{\theta'_{j'},\theta_j}
    \end{pmatrix} \, ,
\end{align}
with entries $[V^\mathrm{F}]_{\theta' \theta}$ given in Eqs.~(\ref{eq:vfinBC}) and (\ref{eq:vfinCC}).

At last, we note that the helicity representation becomes ill-defined for a spin-$s$ particle ($s\neq0$) with zero momentum. 
In most physical systems of practical interest, the coefficient matrix $X$ for such a case can be derived based on the reduction of the restriction representation $O(3) \to G$.
The detailed derivation is shown in the Supplemental Material.

The $\omega$ is conventionally identified as the ground-state isoscalar vector meson, composed of a quark-antiquark pair. 
The $\omega$ couples strongly with $\rho\pi$, despite the $\rho\pi$ threshold lying above the $\omega$ mass. 
Its dominant decay mode is $3\pi$, while the dominant decay mode of the $\rho$ is $2\pi$.
We therefore introduce bare $\omega_0$ and $\rho_0$ states, which may be thought of as quark model states, coupled to the $\rho_0\pi$ and $\pi\pi$ channels, respectively. 
The $\pi\pi\pi$ state is also necessary for coupling to the $\rho_0\pi$ channel.
In principle, the $\rho_0$ can also couple to $K\bar{K}$ channel, and $\rho_0\pi / \pi\pi\pi$ can couple to $\phi$, which itself couples to $K\bar{K}$ channel. 
However, we neglect these channels as their energy thresholds are much higher than the mass region of $\omega$ and $\rho$ mesons. 

The corresponding potentials are given by,
\begin{align}
&\bra{\rho_0\pi(\vec{k},\sigma';-\vec{k})} \hat{V} \ket{\omega_0,\sigma}  
\nonumber\\
& = 
\frac{ -ig_{\omega\rho\pi}}{6\sqrt{2}\pi} C^{1\sigma}_{1\sigma-\sigma^\prime;1\sigma^\prime} \mathcal{Y}_{1,\sigma-\sigma^\prime}(\vec{k}) \mathcal{J}^{\omega_0}_{\rho_0\pi}(0;k,k)\label{eq:Vomegarhopi}
\\
&\bra{\rho_0 \pi(\vec{k}',\sigma^\prime;-\vec{k}')} \hat{V} \ket{\rho_0 \pi(\vec{k},\sigma;-\vec{k})}
\nonumber\\
&=\frac{1}{4}\, c_1(\vec{k}'\cdot\vec{k})\delta_{\sigma\sigma^\prime}\mathcal{J}_{\rho_0\pi\rho_0\pi}(k',k',k,k)
\label{eq:Vrhopirhopi}
\\
&\bra{\rho_0,\sigma} \hat{V}  \ket{2\pi(\vec{k}_1;\vec{k}_2)} \equiv v(\vec{k}_1,\vec{k}_2) Y_{1\sigma}(\vec{k}_1^*) 
\nonumber \\
&= \frac{g_{\rho\pi\pi}}{\sqrt{6}\pi} \mathcal{Y}^*_{1\sigma}(\vec{k}_1^*) \mathcal{J}_{\rho_0\pi\pi}(|\vec{k}_1+\vec{k}_2|,k_1,k_2)
\label{eq:Vrhopipi}
\\
&\bra{\rho_0\pi(\vec{k}',\sigma;-\vec{k}')} \hat{V}  \ket{3\pi(\vec{k}_1;\vec{k}_2;\vec{k}_3)} 
\nonumber\\
&=   \sum\limits_{(i,j,l)} \delta^3(\vec{k}'+\vec{k}_i ) \bra{\rho_0(\vec{k}',\sigma)} \hat{V} \ket{2\pi(\vec{k}_j,\vec{k}_l)}
\label{eq:V3pirhopi}
\end{align}
where the kinematic factor $\mathcal{J}^{\alpha_1,\cdots}_{\beta_1,\cdots}\left(p_1,\cdots;k_1,\cdots\right) := \sqrt{\frac{\omega_{\alpha_1}(p_1)\cdots}{\omega_{\beta_1}(k_1)\cdots}}$ arises from non-relativistic normalization of the states. 
$C^{jm}_{j_1m_1,j_2m_2}$ is the Clebsch-Gordan coefficient. 
$\vec{k}^*(\vec{k}_1,\vec{k}_2)$ is the momentum of the first pion in the rest frame of $\vec{k}_1+\vec{k}_2$. $Y_{lm}(\vec{x})$ is the spherical harmonics and $\mathcal{Y}_{lm}(\vec{x})=x^l Y_{lm}(\vec{x})$. 
The sum over $(i,j,l)$ accounts for all possible assignments of the spectator pion in the three-pion state, ensuring Bose symmetry.
The $Z$-diagram is 
dynamically generated by the $\rho_0\pi \to 3\pi \to \rho_0\pi$ transition, ensuring the energy independence of $\hat{H}^\mathrm{F}$. 
To ensure the ultraviolet convergence of the loop integral in the scattering equation, we introduce the following regulating form factors to the interactions in Eqs.~(\ref{eq:Vomegarhopi}-\ref{eq:Vrhopipi}), respectively: $F(k,
\Lambda_{\omega_0})$, $F(k',\Lambda_{\rho_0\pi}) F(k,\Lambda_{\rho_0\pi})$ and $F(k_j^*,\Lambda_{\pi\pi})$ with $F(k,\Lambda)=(\Lambda^2/(k^2 + \Lambda^2))^2$. 
The values of $\Lambda$ are fixed prior to fitting as $\Lambda_{\omega_0}=1.0$ GeV,$\Lambda_{\pi\pi}=\ 1.0$ GeV, $\Lambda_{\rho_0\pi}=\ 0.5$ GeV, respectively.
Varying these values by $\pm 200$ MeV, we find that the resulting pole positions remain stable, while the regulator-dependent bare parameters adjust accordingly to continue to accurately describe the observable lattice QCD spectra.

In constraining the NPHF to lattice QCD data, five free parameters are introduced: three bare coupling constants ($g_{\omega\rho\pi}$, $c_1$, $g_{\rho\pi\pi}$) and two bare masses ($m_{\rho_0}$, $m_{\omega_0}$). 
While the two bare masses are $m_\pi$-dependent, we follow Ref.~\cite{Yan:2024gwp} and assume that the mass splitting $\delta m = m_{\omega_0}- m_{\rho_0}$ is approximately independent of  $m_\pi$. 
These parameters are determined by performing a simultaneous fit to the lattice spectra at two different values of $m_\pi$. 

Here, we adopt two different schemes to assess the robustness of the extracted parameters under different assumptions. 
For scheme A, we set $\delta m=0$. 
Noting that the bare masses are allowed to be independently fitted at each value of $m_\pi$.
In this case, the number of free parameters is $N_\text{para}=5$. 
%
As we will see, the contribution from $\rho_0\pi \to \rho_0\pi$ is found to be small and $c_1$ is consistent with zero within uncertainties. 
Thus, we introduce the scheme B by setting $c_1 =0$ and fixing $\delta m$ at various values, such that the $N_\text{para}=4$.

\begin{figure}
    \centering
    \includegraphics[width=0.95\linewidth]{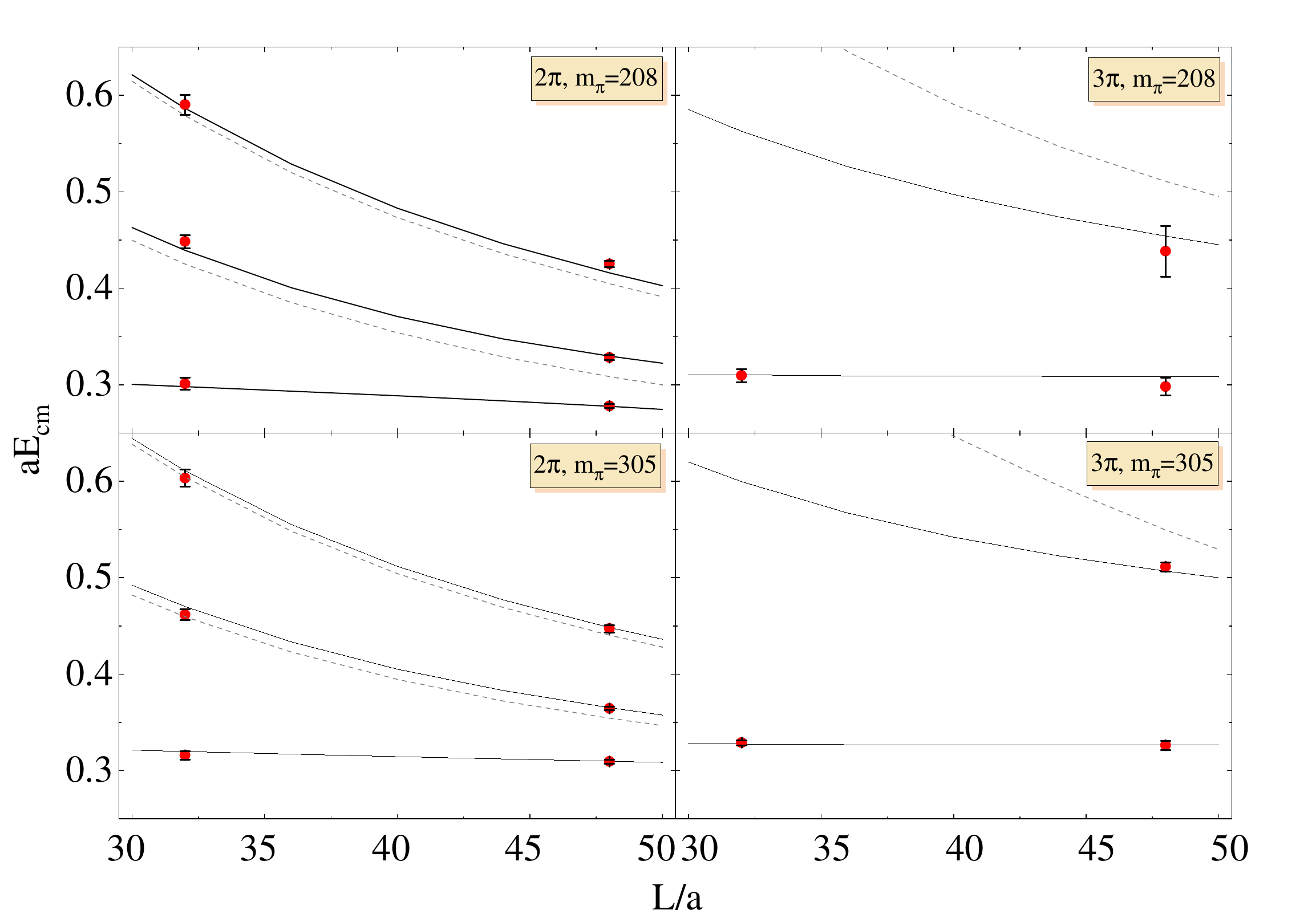}
    \caption{The volume dependence of energy eigenvalues in the $\rho$ and $\omega$ channels at two different pion masses.
    Lattice spectra (red markers) provided in Ref.~\cite{Yan:2024gwp} are used to constrain the eigenvalues $E_n(L)$ (solid lines) of the finite-volume Hamiltonian with parameters determined in scheme $B_3$. The dashed lines represent the $2\pi$ or $3\pi$ non-interacting energy levels. 
    The upper and lower rows correspond to $m_\pi=208$ and $305$ MeV, respectively. 
    The left and right columns refer to $2\pi$ and $3\pi$ spectra, respectively. }
    \label{fig:spectra}
\end{figure}

The fit results are displayed in Table~\ref{tab:fitting result of parameters} and Fig.~\ref{fig:spectra}. 
The masses and couplings for the $\rho$ sector are well constrained and remain stable across both schemes, as they are largely determined by the spectrum of $2\pi$ system . 
In scheme B, we vary the fixed $\delta m$ and the results show a strong correlation between $g_{\omega\rho\pi}$ and $\delta m$. 
This behavior is expected since $g_{\omega\rho\pi}$ controls the strength of the self-energy correction to the bare $\omega_0$ and can partially compensate changes in the assumed $\delta m$.

\begin{table}[tbp]
    \caption{Fit parameters obtained from constraining the NPHF to lattice QCD results. 
    $\hat{\chi}^2$ denotes the reduced chi-squared $\chi^2/\text{d.o.f}$.
    For schemes B1 to B3, we fix $\delta m = 0,\ 15,$ and $30$ MeV, respectively.
    All quantities are given in units with the appropriate powers of $\text{[MeV]}$.  
    The regulators $\Lambda_{\omega_0},\,\Lambda_{\pi\pi}$ and $\Lambda_{\rho_0\pi}$ are fixed at $1.0,\ 1.0,$ and $0.5$ GeV, respectively.}
    \label{tab:fitting result of parameters}
    \begin{center}
    \begin{ruledtabular} 
    \renewcommand{\arraystretch}{1.5} 
    \begin{tabular}{ccccccc}
          & $\hat{\chi}^2$ & $m_{\rho_0}^{208}$ & $m_{\rho_0}^{305}$ & $g_{\rho\pi\pi}\,$ &  $g_{\omega\rho\pi}$ &  $c_1/10^{-5}$  \\ 
        \hline 
        A & $1.2$ & $831(13)$ & $872(10)$ & $7.70(0.45)$ & $29.0^{+12.2}_{-5.4}$ & $1.3^{+4.4}_{-1.5}$ \\ 
        B1 & $1.2$ & $831(13)$ & $872(10)$ & $7.67(0.45)$ &  $26.4(3.5)$ & $0$(fixed) \\ 
        B2 & $1.1$ & $831(13)$ & $872(10)$ & $7.69(0.45)$ & $30.7(3.1)$ & $0$(fixed) \\ 
        B3 & $1.1$ & $832(13)$ & $872(10)$ & $7.70(0.44)$ & $34.4(3.0)$ & $0$(fixed) \\
    \end{tabular}
    \end{ruledtabular}
    \end{center}
\end{table}

We then extract the pole positions of the $\rho$ and $\omega$ in infinite volume using the potentials constrained by the finite-volume lattice spectrum.
The pole position of the $\rho$ meson, $z_\rho$ is the solution of $z-{m_{\rho_0}}-\Sigma_{\pi\pi}(0;z)=0$ on the second Riemann sheet, where $\Sigma_{\pi\pi}(0;z)$ denotes the self-energy for the $\pi\pi$ loop in the $\rho$ rest frame. 
The pole position of the $\omega$ meson, $z_\omega$, is determined by the pole of the off-shell $T$-matrix of $\rho_0\pi\to\rho_0\pi$ as follows, 
\begin{align}
    T(p,k;z) = & U(p,k;z) + 
    \nonumber \\ 
    \int_0^\infty q^2 dq&\,\, U(p,q;z)\, {G}_2(q;z+i0^+)\, T(q,k;z)\, ,
    \label{eq:Tmat of omega}
\end{align}
where the effective potential ${U}(p,k;z)$ contains the contact term, the $s$-channel $\omega_0$ exchange and the dynamically generated $Z$-diagram, as defined in Eqs.~(\ref{eq:Vomegarhopi}-\ref{eq:V3pirhopi}). 
The dressed $\rho\pi$ propagator $G_2(q;z)$ incorporates $\Sigma_{\pi\pi}(q;z)$.
Explicit expressions for $U$, $G_2$ and $\Sigma_{\pi\pi}$ are provided in the Supplemental Material.

With Eq.~(\ref{eq:Tmat of omega}), the pole position $z_\omega$ is determined by the condition 
\begin{align}
    \det ( \delta_{ij} - U(p_i,p_j;z_\omega)\, G_2(p_j;z_\omega)\, p_j^2 \, w_j) = 0\, ,
    \label{eq:det:real}
\end{align}
where ($p_j$,$w_j$) are the quadrature points and the weights along a proper integration path.
Identifying a suitable integration path in momentum space is nontrivial because both the $Z$-diagram contribution in $U$ and the self-energy $\Sigma_{\pi\pi}$ in $G_2$ generate a unitary cut along the real axis starting at the threshold of $3\pi$. 
To extract a resonance pole, analytic continuation to the second Riemann sheet is required. 
The integration contours in Eq.~(\ref{eq:Tmat of omega}) and $\Sigma_{\pi\pi}$ are simultaneously rotated from the positive real axis to the ray $\text{arg} = -\theta$ (denoted by $\mathcal{C}_\theta$), {\it i.e.}, $p_i\to p_i\, e^{-i\theta}$ in Eq.~(\ref{eq:det:real}). 
Consequently, the singular region associated with on-shell intermediate $3\pi$ states and the branch cut related to the $\rho$-resonance (defined by the pole of $G_2(qe^{-i\theta}; z)$) are well separated from the vicinity of the $\omega$ pole region.
Typically, $0<\theta<\pi/4$ are numerically stable and here we choose $\theta=\pi/6$. Details can be found in the Supplemental Material. 
The resulting values of $z_\rho$ and $z_\omega$ are summarized in Table~\ref{tab:poles}. 
\begin{table}[tbp]
    \centering
    \caption{Numerical values for the $\rho$- and $\omega$-meson poles, $z_\rho$ and $z_\omega$ respectively, obtained from fits to finite-volume lattice QCD data, and brought to the infinite volume via the NPHF.
    At $m_\pi=305$ MeV, $z_\omega$ is found on the real axis below threshold in the first Riemann sheet, corresponding to a bound state. 
    The remaining poles reported for each fitting schemes are found on the second Riemann sheet, corresponding to resonance states. 
    The stability of the pole locations across different schemes suggests a weak dependence on the modeling assumptions.}
    \label{tab:poles}
    
    \begin{ruledtabular}
    \renewcommand{\arraystretch}{1.4} 
    \begin{tabular}{cccc} 
        $m_\pi$ & scheme & $z_\rho$ & $z_\omega$ \\ 
        \hline
        \multirow{4}{*}{$208$} 
          & A  & $742.3(5.3)-i55.0(5.3)$  &  $777.9(14.0) - i0.31(0.09)$   \\
          & B1 & $744.8(5.5)-i55.3(5.6)$ &  $779.0 (10.0) - i0.23(0.07)$  \\
          & B2 & $743.1(5.2)-i54.9(5.1)$  &  $ 776.7(10.2) - i0.30(0.09)$ \\
          & B3 & $743.3(5.3)-i55.1(5.4)$  &   $776.3( 9.7) - i0.33(0.07)$ \\
        \hline 
        \multirow{4}{*}{$305$} 
          & A  & $790.8(4.8)-i28.1(3.3)$ & $823.3(5.8)$  \\
          & B1 & $792.0(4.8)-i27.7(3.3)$ &  $829.6(5.1)$  \\
          & B2 & $791.5(5.0)-i28.2(3.4)$ & $827.2(5.4)$  \\
          & B3 & $791.0(4.9)-i28.2(3.5)$ & $826.8(5.3)$  \\ 
    \end{tabular}
    \end{ruledtabular}
\end{table}

In summary, we have extended the FVH formalism~\cite{Yu:2025gzg} to the NPHF to analyze multi-particle systems with isospin symmetry in a finite volume. 
We have provided a systematic procedure for the irreducible representation decomposition of the Hamiltonian, enabling an efficient treatment of coupled multi-channel dynamics.  
As a concrete application, we have performed a comprehensive analysis of the $\omega$-meson at $m_\pi = 208$ and $305$ MeV, based on the lattice QCD spectra from Ref.~\cite{Yan:2024gwp}. 
The Hamiltonian incorporates a bare $\omega_0$ state, the $\rho_0\pi$, and $3\pi$ channels, providing a unified description of single-, two-, and three-body dynamics. 
Through fitting the lattice spectra, we determined the free parameters in the potentials and subsequently used these to extract the $\omega$-pole in infinite volume. 
The results agree with those obtained using the FVU approach~\cite{Mai:2017bge} within uncertainties, demonstrating the consistency between the two formalisms. 

In terms of perspective, the nonperturbative Hamiltonian framework offers a promising tool for studying various three-particle systems, such as the Roper resonance and exotic states such as $T_{cc}$. 
Moreover, similar to some frameworks in studies of nuclear physics on the lattice, the NPHF formalism has the potential for straightforward extensions to $N$-body systems with $N>3$. 
Technical challenges regarding the storage and computation of very large matrices will be an important direction for future developments.

\section*{Acknowledgements}
The authors want to thank the useful discussions with Lu Meng, Feng-kun Guo and Bing-Song Zou. 
This work is supported by the National Natural Science Foundation of China under Grant Nos. 12221005, 12547111 and 12275046,
and by the Chinese Academy of Sciences under Grant No. YSBR-101.
This research is supported by the University of Adelaide and by the Australian Research Council through Discovery Projects DP210103706 (DBL) and DP230101791 (AWT).
%
%

\bibliography{ref}


\clearpage
\onecolumngrid
\appendix

\section*{Supplemental Material}

\section{Details in helicity representation}

The helicity state is defined as 
\begin{align}
     \ket{\{\vec{n},\lambda\}_N}   = \left(\prod\limits_{i=1}^N\sum\limits_{\sigma_i} D^{s_i}_{\sigma_i\lambda_i}(R_\text{st}(\vec{n}_i)) \right) \ket{\{\vec{n},\sigma\}_N}
\end{align}
where $s_i$, $\sigma_i$ and $\lambda_i$ are the spin, the canonical polarization, and the helicity of the $i$-th particle, respectively. 
$R_{\text{st}}(\vec{n})=e^{-i\theta_{\hat{n}} J_z}\, e^{-i\phi_{\hat{n}} J_y}$ defines the standard rotation, and $D^{J}$ is the Wigner-D matrix. 
For potentially half-integer spin, we adopt the convention $0\leq\theta_{\hat{n}}\leq\pi$ and $-\pi\leq\phi_{\hat{n}}<\pi$. 
When $\vec{n}\parallel \hat{e_z}$ we define $\phi_{\hat{n}} = 0$ if $\vec{n}\cdot\hat{e_z}>0$ and $-\pi$ otherwise. 
For any $A\subset G$ we denote $A^+$ and $A^-$ as two complementary sets consist of proper and improper rotations, respectively. 
The transformation rule of helicity states under $G$ is 
\begin{align}
g\ket{\{\vec{n},{\lambda}\}_N} &= \exp\left(-i\sum\limits_{i=1}^N \lambda_i\, \varphi_w(\vec{n}_i,g) \right)\ket{\{g\vec{n},\lambda\}_N} \quad\quad  \forall g\in G^+
\\
\mathbb{P}\,\ket{\{\vec{n},{\lambda}\}_N} &= \eta \, \exp\left(i\pi\sum\limits_{i=1}^N s_i\, \Delta(\phi_{\hat{n}_i}) \right) \ket{\{-\vec{n},-{\lambda}\}_N} ,
\end{align}
where $\eta$ denotes the intrinsic parity and $\Delta(\phi)=-1$ if $0\leq\phi<\pi$ while $1$ if $-\pi\leq\phi<0$. 
$\varphi_w(\vec{n},g)$ is the Wigner angle defined by the equation $D^\frac{1}{2}\left(e^{-i\varphi_w(\vec{n},g)J_z}\right) = D^\frac{1}{2}\left(R^{-1}_{\text{st}}(g\vec{n})\, g \, R_{\text{st}}(\vec{n})\right)$, and it furnishes a one-dimensional representation of $\LG(\vec{n})$, namely,  $\varphi_w(h_1h_2,\vec{n})=\varphi_w(h_1,\vec{n})+\varphi_w(h_2,\vec{n})$ for any $h_1,h_2\in\LG(\vec{n})$.

Given $\{\vec{n}\}^r$, the sum range of $\{\lambda\}$ depends on $\Per(\{\vec{n}^r\})$. 
To clarify, we define a map $\Phi:\mathrm{Per}(\{\vec{n}\})\to S_n$ such that $\Phi(g)(\{\vec{n}\})=g\{\vec{n}\}$. (Note that it is $g\to(\Phi(g))^{-1}$ a homomorphism.) 
For a simple example, $\Phi(\LG(\{\vec{n}\}))=\mathbb{I}$. 
For any $g_0\in \mathrm{Per}(\{\vec{n}\})$,
\begin{align}
    \mathcal{S}(\{\vec{n}^r,\lambda\}_N) &=  \text{span}\left\{ \hat{g}\hat{g}_0\ket{ \alpha_a(\{\vec{n}^r,\,\lambda\}_N) } \,|\,g \in G\,,a=1,\cdots\right\}
    \nonumber \\
    &= \text{span}\left\{ \hat{g}\ket{ \alpha_a(\{\vec{n}^r,\,\Phi(g_0)^{-1}\mathcal{P}(g_0)\lambda \}_N} \,|\,g \in G\,,a=1,\cdots\right\} 
    = \mathcal{S}(\{\vec{n}^r,\Phi(g_0)^{-1}\mathcal{P}(g_0)\lambda\}_N)\, ,
\end{align}
where $\mathcal{P}(g_0)$ is the parity of $g_0$.
Consequently, the range of $\{\lambda\}$ is given by the quotient set $\{\lambda_i=-s,\cdots,s\}/\sim$ where $\{\lambda\}\sim \Phi(g)^{-1}\mathcal{P}(g)\{\lambda\}\,,\forall g\in \mathrm{Per}(\{\vec{n}^r\})$. 
Such equivalence relations are non-trivial when $\Per(\{\vec{n}^r\})/(\LG(\{\vec{n}^r\})^+)$ is non-empty.


We now discuss the treatment of states where certain particles possess zero momentum. In spin space, zero-momentum particles must revert to the canonical polarization representation, whereas the other particles remain in the helicity representation. Without loss of generality, we assume the first $M$ ($M \leq N$) particles have zero momentum. To distinguish these $M$ particles from the remaining $N-M$ particles, the permutation symmetry in isospin space is reduced from $(S_N)_{\text{iso}}$ to its subgroup $(S_M \times S_{N-M})_{\text{iso}}$. We now first construct states with definite total isospin $(I,I_z)$ and furnish the irrep $[\kappa_M]_\text{iso} \times [\kappa_{N-M}]_\text{iso}$. The final basis furnishing the irrep $\Gamma$ is obtained by independently constructing the basis for the $\Gamma_M$ and $\Gamma_{N-M}$ of $G$ in the tensor spaces of the first $M$ and the remaining $N-M$ particles, respectively, followed by the tensor product decomposition $\Gamma_M \times \Gamma_{N-M} \to \Gamma$. The construction for the latter $N-M$ particles follows the previously described procedure. For the first $M$ particles, the construction proceeds as follows. First, analogous to the isospin space, we construct states in spin space with definite total spin $(S,S_z)$ and $[\kappa_M]_\text{spin}$ of $(S_M)_{\text{spin}}$. Subsequently, taking the tensor product of isospin and spin spaces, the physical states need to be either totally symmetric or totally antisymmetric under permutation in the combined space. Given the properties of $S_N$ that $[\kappa] \times [\kappa] \to [N]$ and $[\kappa] \times [\kappa]^* \to [1^N]$, $[\kappa_M]_{\text{spin}}$ can only be either $[\kappa_M]_{\text{iso}}$ or $[\kappa_M]_{\text{iso}}^*$. Finally, the states satisfying the correct statistical properties will furnish representation $D^S$. Utilizing the reduction coefficients of the restriction $D^S \to \Gamma_M$, we obtain the basis furnishing $\Gamma_M$ and the overall construction is then completed.

\section{ Linear combination coefficients of $\omega\to 3\pi$ }

We present the linear coefficients for the $\Gamma=T_{1}^{-}$ irrep, denoting the representation matrix simply as $D \equiv D^{T_{1}^{-}}$. 

For $\omega_0$, $[X]$ is a $3\times1$ matrix with entries $[X]_{\sigma,r=1}=\sqrt{\frac{1}{2}}\left(-\delta_{\sigma,1} + \delta_{\sigma,-1}\right)$. 

For $\rho_0\pi$, $[X]$ is identical to the $\omega_0$ case if the momentum is zero. Otherwise, we must first convert from the polarization representation to the helicity representation. 
The invariant subspace is characterized by $\xi=(\{\vec{n}^r\},\lambda)$. 
The helicity $\lambda$ runs over $\{1,0,-1\}$ if $|\vec{n}^r_{x}|,\ |\vec{n}^r_{y}|,$ and $|\vec{n}^r_{z}|$ are non-zero and different from each other or otherwise $\{1,0\}$ since then $\LG(\vec{n}^r)^-$ is non-empty.
The overlap matrix in Eq.~(\ref{eq:Imat}) becomes
\begin{align}
    [I^\xi] =  \begin{cases}
    \sum\limits_{g\in \LG(\vec{n}^r)^+} 
    {D}^{*}(g) \cdot \left(\mathbb{I}-D^*(h)\right) \quad \forall h\in \text{LG}({\vec{n}^r})^-   & \text{if}\quad {\lambda}=0 \, ,
    \\
    \sum\limits_{g\in \LG(\vec{n}^r)^+} e^{-i{\lambda}\varphi_w(\vec{n}^r,g)} {D}^{*}(g) & \text{if}\quad {\lambda}\neq0 \, .
    \end{cases}
\end{align}
The row index $(\vec{n}, \sigma)$ of $[X]$ can equivalently be specified by the coset element $g \in G/\mathrm{LG}(\{\vec{n}\})$, due to the one-to-one correspondence $(g,\sigma) \leftrightarrow (g\vec{n}^r,\sigma)$. 
The specific expression is given by 
\begin{align}
[X_\xi]_{\vec{n}\,\sigma,\,r} = [X]_{g\,\sigma,\,r} =  &\frac{1}{4}\, D^{s=1}_{\sigma,\mathcal{P}(g)\lambda}(R_\text{st}(\vec{n}))\,\mathcal{P}(g)\times \notag\\
&\begin{cases}
|\LG(\vec{n}^r)|^\frac{1}{2}\sum\limits_{\nu}   D^*_{1\nu}(g)(c^r)_\nu  & \text{if}\quad \lambda=0 \, ,
\\
|\LG(\vec{n}^r)^+|^\frac{1}{2} \sum\limits_{\nu' \nu} D^*_{1 \nu'}(g) \left( \delta_{\nu'\nu} - e^{-i\varphi_w(\vec{n}^r,g)} D^*_{\nu'\nu}(h) \right) (c^r)_\nu \quad\forall h\in \text{LG}({\vec{n}^r})^-
& \text{if}\quad \lambda \neq 0 \, .
\end{cases}
\end{align}

For the isoscalar $3\pi$ system, irrep of $S_3$ can only be the one-dimensional antisymmetric one. Therefore, the momenta of the three pions must be distinct. The invariant subspace is characterized by $\xi=\{\vec{n}^r\}$ and the index of $[X]$ is $\{\vec{n}\}$.
Firstly, the $I$-matrix in Eq.~(\ref{eq:Imat}) becomes
\begin{align}
 [I^\xi]
   = \sum\limits_{g\in \Per(\{\vec{n}^r\})} \mathcal{P}(g)\, \mathcal{R}^{[1^3]}(\Phi(g))\, D^*(g) \, ,
\end{align}
%
The row index of $[X]$ can also be specified by the left-coset element. Therefore, $[X]$ reads
\begin{align}
    [X_\xi]_{\{\vec{n}\},r} = [X]_{g,r} = \frac{1}{4}\,|\Per(\{\vec{n}^r\})|^\frac{1}{2}\,\sum\limits_{\nu}  D^{*}_{1\nu}(g) (c^r)_{\nu}\,\mathcal{P}(g)
\end{align}

\section{Details for the determination of the $\omega$-pole}

The pole position of the $\pi\pi\pi \to \pi\pi\pi$ scattering amplitude is determined from the singularity of the $P$-wave off-shell $T$-matrix for $\rho_0\pi \to \rho_0\pi$~\cite{Doring:2018xxx}, which is obtained by solving 
Eq.~(\ref{eq:Tmat of omega}).
The dressed $\rho\pi$ propagator $G_2(q;z)$ reads
\begin{align}
     {G_2}(p;z) &=\left( z - \omega_{\rho_0}(p) - \omega_\pi(p) - \Sigma_{\pi\pi}(p;z)  \right)^{-1} \, ,
     \\
     \Sigma_{\pi\pi}(p;z) &= \int q^{2}dq \frac{m_{\rho_0}}{\omega_{\rho_0}(p)}\frac{2\omega_\pi(q)}{\sqrt{ p^2 + 4\omega^2_\pi(q)}}\frac{\frac{1}{2}\left(v(\vec{q},-\vec{q})\right)^2 }{z-\omega_\pi(p)-\sqrt{ p^2 + 4\omega^2_\pi(q)} + i0^+} \,.
     \label{eq:Sigmapipi}
\end{align}
The effective $\rho_0\pi \to \rho_0\pi$ interaction is denoted as $U(p,k;z)=C(p,k)+R(p,k;z)+Z(p,k;z)$ where the contact potential $C(p,k)$, the s-channel $\omega_0$ exchange potential $R(p,k;z)$, and the $Z$-diagram contribution $Z(p,k;z)$ are expressed as
\begin{align}
C(p,k) &= \frac{\pi c_1}{3} \left(p\,{F}(p,\Lambda_{\rho_0\pi})\right)\left(k\,{F}(k,\Lambda_{\rho_0\pi})\right) \mathcal{J}_{\rho_0\pi \rho_0\pi}(p,p,k,k) \, ,
\\
R(p,k;z) &= \frac{1}{z-m_{\omega_0}}\left(\frac{g_{\omega\rho\pi}}{6\sqrt{2}\pi}\right)^2   \left(p\,{F}(p,\Lambda_{\omega_0})\right)\left(k\,{F}(k,\Lambda_{\omega_0})\right) \mathcal{J}^{\omega_0}_{\rho_0\pi}(0;k,k)\mathcal{J}^{\omega_0}_{\rho_0\pi}(0;p,p) \, ,
\\
Z(p,k;z) &= 4\pi^{3/2}\sum\limits_{\sigma,\sigma^\prime}C_{1\sigma-\sigma^\prime;1\sigma^\prime}^{1\sigma} C_{10;1\sigma}^{1\sigma}N_{1,\sigma-\sigma^\prime} \int dx\, \frac{P_{1,\sigma-\sigma^\prime}(x)\,v_{\sigma^\prime}^*(-\vec{k},\vec{p}+\vec{k})\,v_\sigma(\vec{p}+\vec{k},-\vec{p})}{z-\omega_\pi(k)-\omega_\pi(p)-\omega_\pi(|\vec{p}+\vec{k}|)+i0^+} \,.
\label{eq:Zdiagram}
\end{align}
Here, $N_{lm}:=\sqrt{\frac{(2l+1)}{4\pi}\frac{(l+m)!}{(l-m)!}}$, $v_\sigma(\vec{k}_1,\vec{k}_2):=\int d\hat{k}^*\,Y_{1\sigma}^*(\vec{k}^*)\bra{\rho_0,\sigma} \hat{V}  \ket{2\pi(\vec{k}_1;\vec{k}_2)}$ and the integration variable $x$ denotes $\hat{p}\cdot\hat{k}$. 
Due to the presence of the resonant term, $R(p,k;z)$, and assuming no other dynamically generated poles, the singularity of $T(p,k;z)$ originates from the pole of the fully dressed $\omega$ propagator, which is identified as the $\omega$-pole. 
Eq.~(\ref{eq:Tmat of omega}) can be rewritten in operator form, which yields a formal solution,
\begin{align}
\hat{T}(z)=\left(1-\hat{U}(z)\hat{{G}}_2(z)\right)^{-1}\hat{U}(z),
\end{align}
so that the pole position, $z_\omega$, can be solved by 
\begin{align}
\det\left(1-\hat{U}(z)\hat{{G}}_2(z)\right)=0 \, . 
\end{align}
This is equivalent to the condition that the operator $UG_2$ possesses at least one eigenvalue equal to 1.

Due to the loop integrals in Eq.~(\ref{eq:Sigmapipi}) and Eq.~(\ref{eq:Zdiagram}), $U G_2$ grows a unitary branch cut starting from $3m_\pi$ on the real-axis in the complex $z$-plane. 
To obtain a $\omega$-resonance pole located on the second Riemann sheet, Eq.~(\ref{eq:det:real}) should be analytically continued by, for example, the contour deformation method: 
\begin{align}
    \int_{\mathcal{C_\theta}} q_\theta^2\, dq_\theta  \,U(p_\theta,q_\theta;z)\, {G}_2(q_\theta;z)\, \psi(q_\theta;z) 
   =  \psi(p_\theta;z) \, ,
   \label{eq:rotated UGeigen}
\end{align}
where $\mathcal{C}_\theta$ is the ray $\arg=-\theta$ and the variable with $\theta$ is defined on the ray. 
Note that the integration path in Eq.~(\ref{eq:Sigmapipi}) should also be simultaneously rotated into $\mathcal{C}_\theta$:
\begin{align}
    \Sigma_{\pi\pi}(p_\theta;z) := \int_\mathcal{C_\theta} q^{2}_\theta dq_\theta \frac{m_{\rho_0}}{\omega_{\rho_0}(p)}\frac{2\omega_\pi(q_\theta)}{\sqrt{ p_\theta^2 + 4\omega^2_\pi(q_\theta)}}\frac{\frac{1}{2}\left(v(\vec{q}_\theta,-\vec{q}_\theta)\right)^2 }{z-\omega_\pi(p_\theta)-\sqrt{ p_\theta^2 + 4\omega^2_\pi(q_\theta)} + i0^+} \, .
\end{align}

\begin{figure}[htbp]
    \centering
    \includegraphics[width=0.5\linewidth]{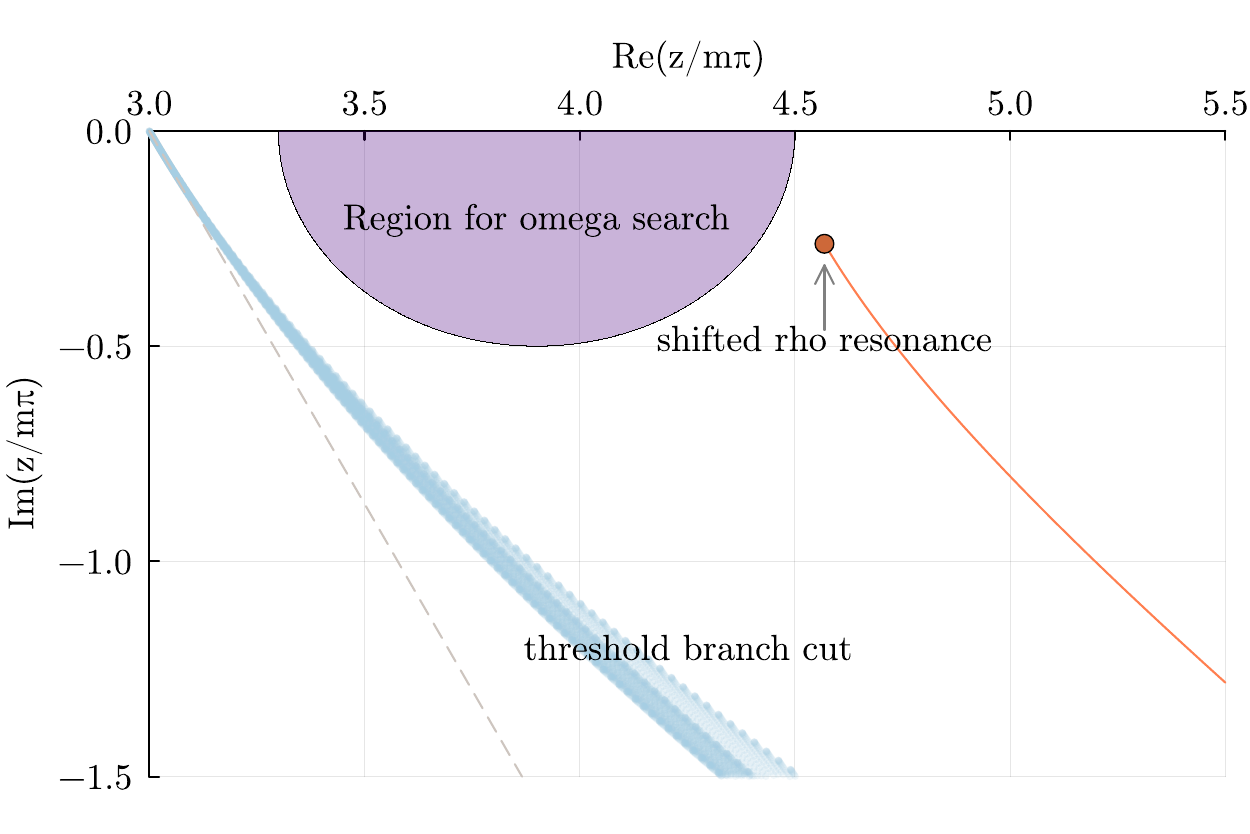}
    \caption{ The lower half of the complex $z$-plane on the second Riemann sheet. 
    The dashed line indicates the rotated $3\pi$ unitary cut, defined by $\text{arg}(z-3m_{\pi})=-2\theta$, when non-relativistic kinematics $\omega_\pi(p)=m_\pi+\frac{p^2}{2m_\pi}$ is adopted. 
    For relativistic kinematics, the cut is distorted into the blue-shaded region. 
    The orange line denotes the branch cut associated with the $\rho$-resonance, ending at $z_\rho + m_\pi$. 
    The purple region indicates the search area for the $\omega$-resonance pole.}
    \label{fig:omega}
\end{figure}

As shown in Fig.~\ref{fig:omega}, the singular region from the opening of the $3\pi$ channel is now rotated into the lower half plane, which can be explicitly parametrized as $z=3 m_{\pi} + e^{-2i\theta}t$ with $t\in\mathbb{R}^+$ if non-relativistic kinematics $\omega_\pi(p)=m_\pi + \frac{p^2}{2m_\pi}$ is adopted. 
For relativistic kinematics the region is distorted.  
In addition, since the second Riemann sheet of $G_2$ is also reached, there is a branch cut associated with the $\rho$-resonance determined by $G_2(pe^{-i\theta};z)=0$ with running $p$. 
The endpoint, namely, when $p=0$, is exactly $z_\rho + m_\pi$. 
Away from the two singular regions, $UG_2$ is analytic along $C_\theta$ and therefore the integral equation Eq.~(\ref{eq:rotated UGeigen}) can be discretized into an algebraic equation using the standard Gauss quadrature:
\begin{align}
    \sum\limits_{j} \,U(p_{\theta i},p_{\theta j};z)\, q_{\theta j}^2\, w_j\, {G}_2(p_{\theta j};z)\, \psi(p_{\theta j};z) 
   =  \psi(p_{\theta i};z) \quad,\forall i \, ,
\end{align}
which finally leads to Eq.~(\ref{eq:det:real}) with $p\to p\, e^{-i\theta}$.

\end{document}